\documentclass[journal]{IEEEtran}
\usepackage{graphicx}

\begin{document}
\title{Calibration System with Optical Fibers for Calorimeters at Future Linear Collider Experiments}

\author{
        Jaroslav Zalesak\\ 
        \emph{for the CALICE Collaboration}
\thanks{Manuscript received November 15, 2011.}
\thanks{J. Zalesak is with the Institute of Physics of the Academy of Science, Na~Slovance 2, CZ-18121 Prague, 
Czech Republic (phone: 420-26605-2707), (\emph{e-mail: zalesak@fzu.cz})}%

\thanks{This work is supported by the Ministry of Education, Youth and Sports of the Czech Republic under 
the projects AVO Z3407391, AVO Z10100502, LC527 and INGO LA09042.}%
}

\maketitle
\pagestyle{empty}
\thispagestyle{empty}

\begin{abstract}

We report on several versions of the calibration and monitoring system dedicated to scintillator tile 
calorimeters built within the CALICE collaboration and intended for future linear collider 
experiments. Whereas the first, a 1 m3 analogue hadron calorimeter prototype, was already built and 
tested in beam, second --- technological prototype --- is currently being developed. Both prototypes are 
based on scintillating tiles that are individually read out by new photodetectors, silicon 
photomultipliers (SiPMs). Since the SiPM response shows a strong dependence on the 
temperature and bias voltage and the SiPM saturates due to the limited number of 
pixels, it needs to be monitored. The monitoring system has to have sufficient flexibility to perform 
several different tasks. The self-calibration property of the SiPMs can be used for the gain monitoring 
using a low intensity of the LED light. A routine monitoring of all SiPMs during test beam operations 
is achieved with a fixed-intensity light pulse. The full SiPM response function is cross-checked by 
varying the light intensity from zero to the saturation level. We concentrate especially on the last 
aspect --- the high dynamic range.  

\end{abstract}

\begin{IEEEkeywords}
ILC, CALICE, Calorimetry, LED, Calibration, Optical fiber, Silicon photomultiplier.
\end{IEEEkeywords}

\section{Introduction}

 \IEEEPARstart{T}{he} CALICE collaboration \cite{Zalesak:CaliceWeb} is developing 
a hadronic calorimeter (HCAL) with very high granularity 
for future linear colliders (ILC, CLIC). The collaboration built 1~m$^{3}$ physics prototype in 
2005--6 \cite{Zalesak:Ahcal} and currently CALICE is building an engineering prototype \cite{Zalesak:Mathias}.

  The HCAL readout chain of these prototypes contains scintillator tiles with embedded 
wavelength-shifting fibers and small SiPM (Silicon Photo Multiplier) photo-detectors. The electrical 
signal is adjusted by a preamplifier and a shaper and digitized by a 12-bit ADC. 

  The variation of characteristics of the complete chain (gain, saturation) depends mainly on changes 
of the temperature and operation voltage of the SiPM. For the correct offline reconstruction of the 
energy deposition, the calibration runs have to be included into the data-taking process, since the 
condition inside the detector can change.

   The HCAL prototypes will be shortly described in section \ref{subsecPPT} and \ref{subsecEngP}. 
The section \ref{subsecCalib} gives information about the calibration procedure we are using.
The calibration systems for 
these prototypes will be described in section \ref{secLEDDrive} with impact on the electronic principles of 
operation, together with a performance results. The section \ref{secFiber} will describe a solution for 
multi-tiles-illumination by a single optical fiber and also gives insight into the ongoing development.

\subsection{Physics Prototype\label{subsecPPT}}

\begin{figure}
\centering
\includegraphics[width=3.0in]{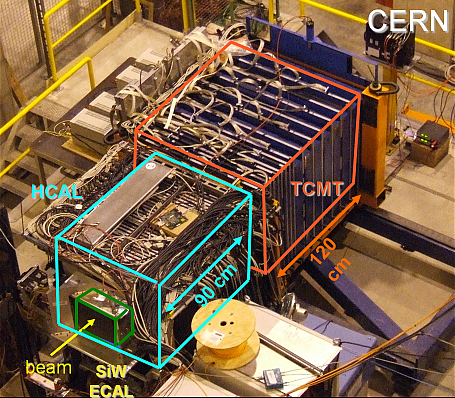}
\caption{Configuration of the ECAL, HCAL and TCMT in the beam test at CERN.}
\label{TbAhcal}
\end{figure}

  The AHCAL (Analogue Hadron CALorimeter) 1 m3 physics prototype with 7608 active readout channels 
has been in test beams at the CERN SPS in 2006 and 2007 (Fig.~\ref{TbAhcal}). 
Beam tests at Fermilab followed in years 2008 and 2009 and now the physics prototype continues 
to run with tungsten absorber at the CERN PS in 2010 and at the SPS in 2011. 

\begin{figure}
\centering
\includegraphics[width=2.97in]{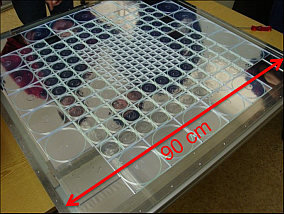}
\caption{Layout of the scintillator tiles inside the cassette, active part of one layer of the AHCAL physics prototype.}
\label{SciLayer}
\end{figure}

  The physics prototype is made of 38 layers with scintillator tiles, interlaced with 16 mm Fe absorber plates
  (or with the 10~mm~W plates). 
The dimension of one layer is 90~cm~x~90~cm and it contains scintillator tiles with different granularity 
(Fig.~\ref{SciLayer}):
20~pcs of 12x12~cm$^2$ tiles, 96~pcs of 6x6~cm$^2$ tiles and first 30~layers have 100~pcs of 
3x3~cm$^2$ tiles in the middle.
Last 8~layers have 6x6~cm$^2$ tiles in the middle instead. The active element of the readout is a 
compact photo-detector --- a silicon photomultiplier SiPM  (see section~\ref{subsecCalib}).
with 1156~pixels. The AHCAL analogue board carries 1~ASIC for 18~SiPM channels. A second board
contains control and configuration electronics and provides correct voltage 
for each SiPM. For the calibration a board called CMB is used (see section~\ref{subsecCMB}).
   
\subsection{Engineering Prototype\label{subsecEngP}}

\begin{figure}[b]
\centering
\includegraphics[width=3.3in]{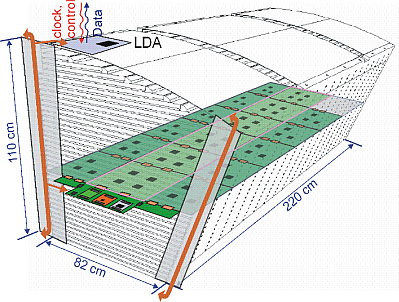}
\caption{Wedge of the HCAL barrel (1/16 of the half-barrel) with one integrated layer shown consisting several HBUs
 and driving and read-out cards.}
\label{AhcalEngineer}
\end{figure}

  Next version of the AHCAL is called the engineering prototype.
  The basic electronics structural element of the engineering prototype (Fig.~\ref{AhcalEngineer}) 
is the HCAL base unit (HBU). 
The engineering prototype is built to demonstrate the feasibility of the electronics integration in the 
HCAL calorimeter for ILC, using 3 cm x 3 cm scintillator tiles with SiPM photo-detectors, embedded readout
\cite{Zalesak:Mark} and implementing a concept of power pulsing.

 Fig.~\ref{AhcalEngineer} shows one half of the HCAL octagon. Each wedge consists of 48 layers. 
Each layer is made of absorber 
(16~mm Fe or 10~mm W) and active layer of 2-3 rows with 6 HBUs connected together in a cassette, 2.2~m~long.

\begin{figure}
\centering
\includegraphics[width=2.4in]{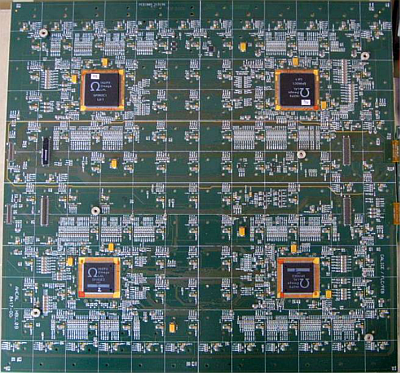}
\caption{The HBU unit with 4 ASICs which read-out 144 scintillator tile channels placed on the back side.}
\label{Hbu}
\end{figure}

  The HBU unit (Fig.~\ref{Hbu}) is 36~cm~x~36~cm large board, which has 144~scintillator tiles 
with embedded SiPMs with 796~pixels on its back side. All 144~channels are read-out by 4~ASICs. 
Details on the HBU are described in~\cite{Zalesak:Mark}. 
Two options of optical calibrations are proposed for the HBU: distributed SMD LEDs~\cite{Zalesak:Mathias}
 and an external LED driver and optical fiber distribution (sections~\ref{secLEDDrive} and~\ref{secFiber}).
  
\subsection{SiPM and Calibration\label{subsecCalib}}

  SiPMs are photo detectors, that consist of hundreds (up to thousands) pixels, each working as an avalanche 
photodiode in Geiger mode. Each pixel is coupled via a resistor to a single output, creating an analogue signal 
with amplitude equivalent to the number of fired pixels. 

\begin{figure}
\centering
\includegraphics[width=1.58in]{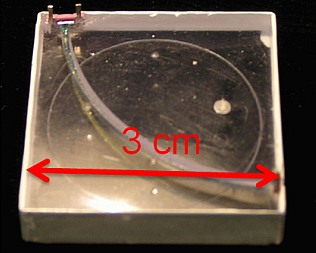}
\includegraphics[width=1.6in]{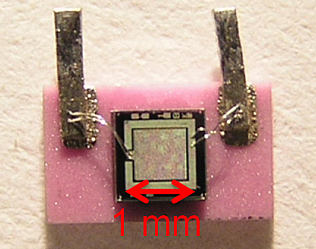}
\caption{(a) Photo of a 3~cm~x~3~cm~scintillator tile  for the physics prototype with the embedded WLS fiber and SiPM;
 (b)~Detail of the SiPM chip.}
\label{SciTileSipM}
\end{figure}

  The light from the scintillator tile (Fig.~\ref{SciTileSipM}(a)) is collected 
by a wavelength-shifting fiber, that re-emits the light in green (where SiPM is most sensitive) and guides 
the light toward the active area of the SiPM (1~mm$^2$, Fig.~\ref{SciTileSipM}(b)). 
The other end of the fiber has a mirror.
  
  The gain of the SiPM varies as a result of the manufacturing process. The typical gain of the SiPM is~10$^6$ 
and is highly dependent on temperature T and bias voltage. The typical gain varies 
by~-1.7~\%/K
and +2.5~\%/0.1V. The temperature and voltage affects also the photodetection efficiency of SiPM, which results
in a response variation of -4.5~\%/K and +7~\%/0.1V in total.  

\begin{figure}
\centering
\includegraphics[width=1.7in]{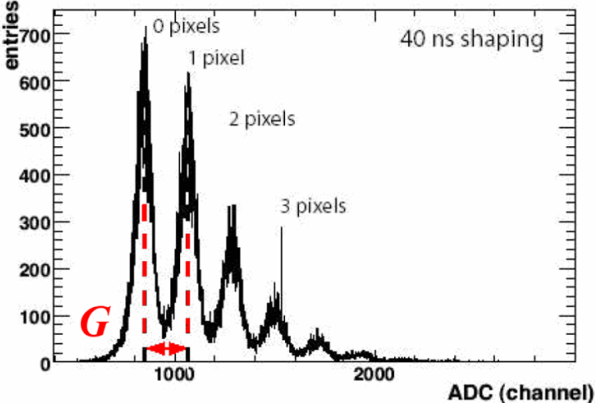}
\includegraphics[width=1.7in]{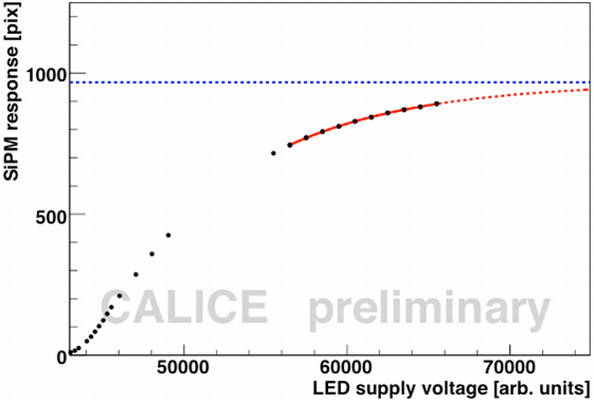}
\caption{(a) Gain extraction from the single photo-electron spectrum; (b) SiPM saturation effects.}
\label{SaturationPhESipM}
\end{figure}

  The gain of the SiPM can be obtained from the Single Photo-electron Spectrum (SPS) as a distance among the peaks 
(Fig.~\ref{SaturationPhESipM}(a)). Low intensity light flashes are required for this calibration measurement. 

  The SiPM needs correction for the saturation curve, as the number of fired pixels is approaching the 
total number of pixels as can be seen in~Fig.~\ref{SaturationPhESipM}(b). 
The saturation curve is also affected by the time structure of the photon shower, as 
the pixels can recover and fire again on the next photon, creating signal effectively higher than the total number 
of pixels in SiPM.

\section{Calibration LED Driver Cards and Principles\label{secLEDDrive}}

  Since the SiPMs are sensitive to changes in temperature and operation voltage, the LED system has been installed 
to monitor the stability of the readout chain in time. The calibration system needs sufficient flexibility to perform 
several different tasks. Gain calibration: we utilize the self-calibration properties of the SiPMs to achieve the calibration 
of an ADC in terms of pixels that is needed for non-linearity corrections and for a direct monitoring of the SiPM gain 
(in~Fig.~\ref{SaturationPhESipM}(a)). Further we monitor all SiPMs during test beam operations with a fixed intensity
light pulse. Saturation: we cross check the full SiPM response function by varying the light intensity from zero to 
the saturation level (in~Fig.~\ref{SaturationPhESipM}(b)).
 
 To satisfy such requirements we have developed several versions of calibration and monitoring boards.
   
       Our calibration boards use an UV-LED as a light source for calibration. The UV-LEDs require a special 
driver in order to make them shine fast ($\leq$10~ns) with an amplitude covering several orders of magnitude 
in light intensity.

\subsection{Calibration and Monitoring Board (CMB)\label{subsecCMB}}

\begin{figure}
\centering
\includegraphics[width=3.5in]{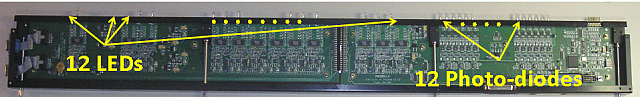}
\caption{Layout of the CMB board: 12 LEDs on the left side and 12 PIN photodiodes on right side.}
\label{CMBPhoto}
\end{figure}

  The first system we developed was called the CMB, Calibration and Monitoring Board 
  \cite{Zalesak:Ivo}. It consists of 12 UV LEDs  (Fig.~\ref{CMBPhoto})  
with a special fast driver optimized for rectangular pulses. The pulses are 10~ns wide in order to match real signals 
from hadron showers in the calorimeter as closely as possible. By varying the control voltage, the LED intensity 
covers the full dynamic range from zero to saturation (about 70 minimum ionizing particles). The LED illuminates 
a fiber-bundle of 18~+~1. One fiber comes back to the CMB to monitor the emitted light by a PIN photo diode 
with preamplifier on the board. The other 18 fibers are rooted to each scintillator tile equipped with one SiPM each.

\begin{figure}
\centering
\includegraphics[width=1.95in]{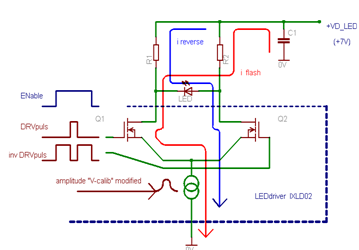}
\includegraphics[width=1.45in]{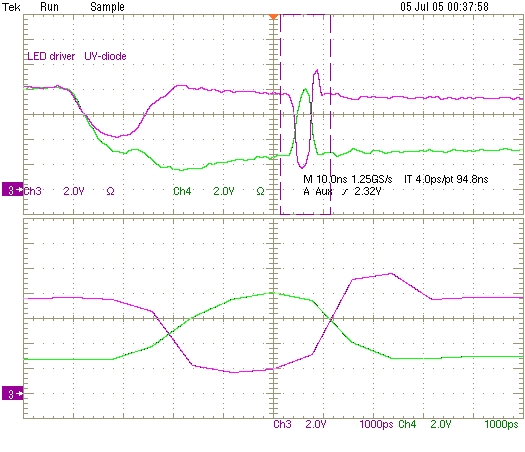}
\caption{(a) Operation principle of the CMB LED driver; (b) Oscillograph from the LED pins during a pulse; 
green: anode, violet: cathode; Bottom part is a zoom of the selected region from the upper part. 
Upper traces: 10~ns per div, bottom traces: 1~ns per div.}
\label{CMBosci}
\end{figure}

  The CMB is controlled by a slow control CAN-bus protocol. The output light is adjustable both in amplitude and 
length (5--100~ns). The trigger is provided externally by T-calib signal. 
The CMB provides readout from temperature sensors from several
places on the detector layer. The amplitude covers the whole range of light intensities, from single photons up to 
the SiPM saturation.

  CMB utilizes a IC from IXIS company (IXLD02). In principle this driver works as a H-bridge, where upper part is 
tightened to the supply voltage through a protective resistor. Bottom part of the H-bridge (two FETs) are 
very fast switches, which are closed in idle (Fig.~\ref{CMBosci}(b), start of the oscillograph) --- no current passes the LED. 
Before the pulse, the LED is reverse biased (controlled by the enable signal in Fig.~\ref{CMBosci}(a)). 
The pulse is generated by reversing the driver's "puls" signals. The LED is biased in the forward direction 
(Fig.~\ref{CMBosci}(b) bottom detail), forcing the LED to emit light. At the end of the pulse, 
the LED is reverse biased again.  The reverse bias is an important step, because without discharging 
the LED tends to continue emitting light. The time of rising and falling edges of the pulse is about~1~ns.

\begin{figure}[!b]
\centering
\includegraphics[width=2.7in]{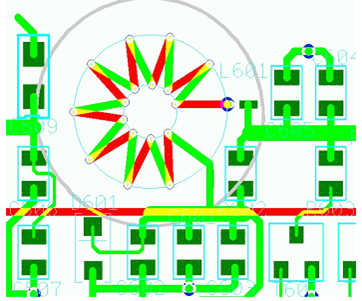}
\caption{Design of a toroidal inductance (for low RFI 35~nH) on a PCB board for the quasi resonant LED driver .}
\label{Toroidal}
\end{figure}

\begin{figure}
\centering
\includegraphics[width=2.1in]{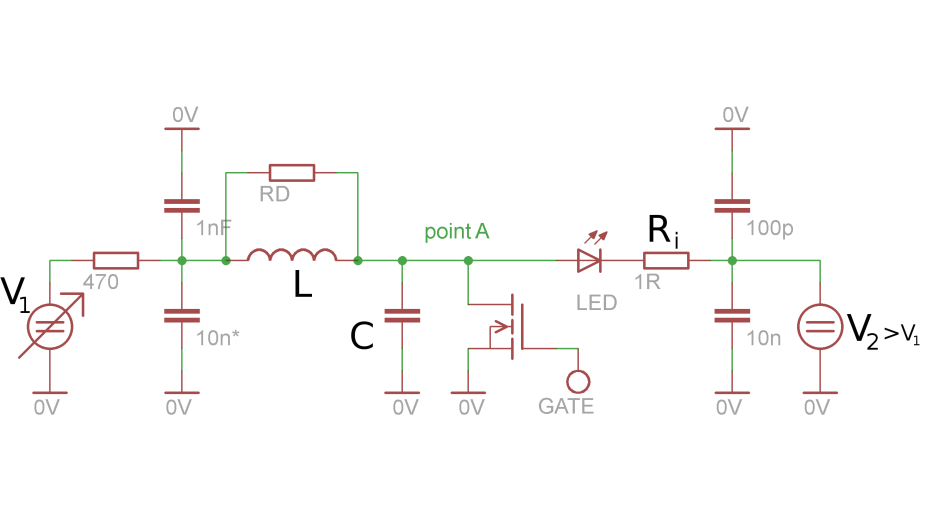}
\includegraphics[width=1.3in]{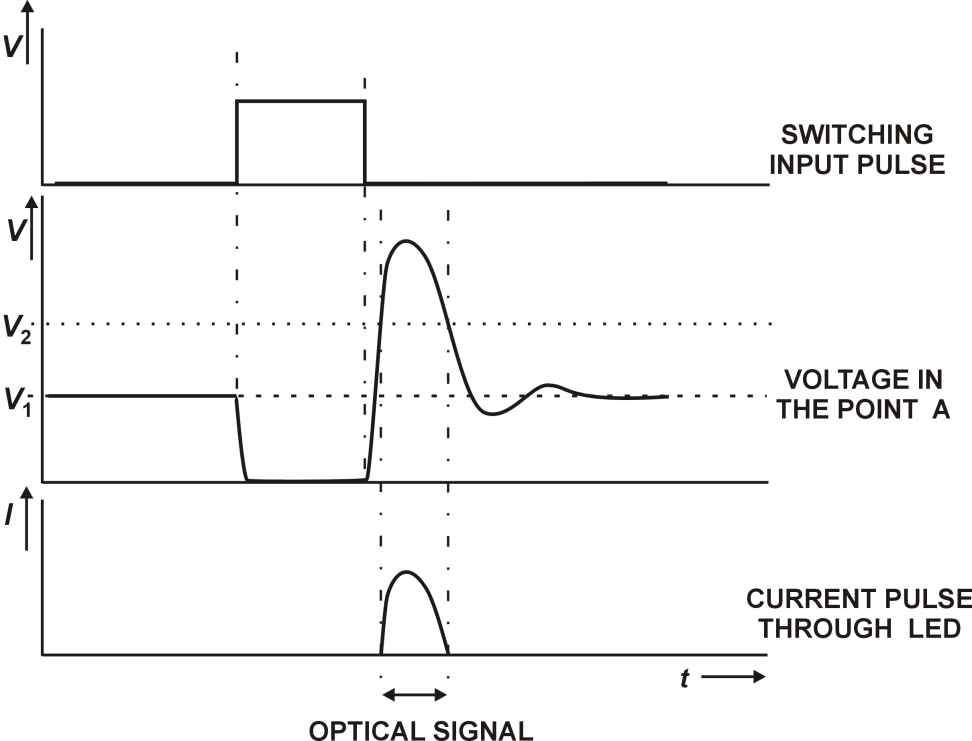}
\caption{Quasi-resonant LED driver: (a) driver circuit; (b) operation principle.}
\label{Quasidriver}
\end{figure}

\subsection{Quasi-resonant LED driver}

 To further improve the performance of the LED driver we abandoned the rectangular pulses. 
The Quasi Resonant LED driver (QRLED) produces short 5~ns long electrical pulses for LEDs of the sinusoidal shape. 
The QRLED is foreseen as a source of the tunable LED light for calibration of SiPMs in the engineering 
hadron calorimeter module. The short electrical pulses are created in the toroidal inductance made directly on the PCB. 
This design resulted from the requirement on the minimal height of the electronic circuitry 
in the compact engineering module. 
 
    The driver uses embedded PCB toroidal inductors (Fig.~\ref{Toroidal}),
 which help to reduce the EMI (electromagnetic interference).

  The LED driver works in principle similarly to the boost DC-DC converter (Fig.~\ref{Quasidriver}). 
In the idle mode, the V2 voltage is higher than V1 and the LED is reverse biased. Before the pulse, 
the inductor (Fig~\ref{Quasidriver}(a):~"L") 
is switched to the ground, enabling the current to flow through the inductor and store energy. When the 
switching transistor is switched off, the circuit is left in the resonance configuration. The energy 
(current through "L") is smoothly transferred to the capacitor (voltage on "C''), back and so on. 
The resonance is heavily dumped by the dump resistor (''RD''), therefore only the first overshoot on "C" 
is high enough to bias the LED directly.

  The first overshoot (sine wave) can be tens of volts, forcing the LED to pass through a current up to 1 A. 
The negative part of the sine wave helps to reverse bias the LED in a very short time. The circuit is tuned such 
that the following sine wave will not overcome the direct-bias threshold and the LED is kept reverse biased as can be seen
in~Fig.~\ref{Quasidriver}(b).

\subsection{External Calibration System --- QMB6}

\begin{figure}[!b]
\centering
\includegraphics[width=1.7in]{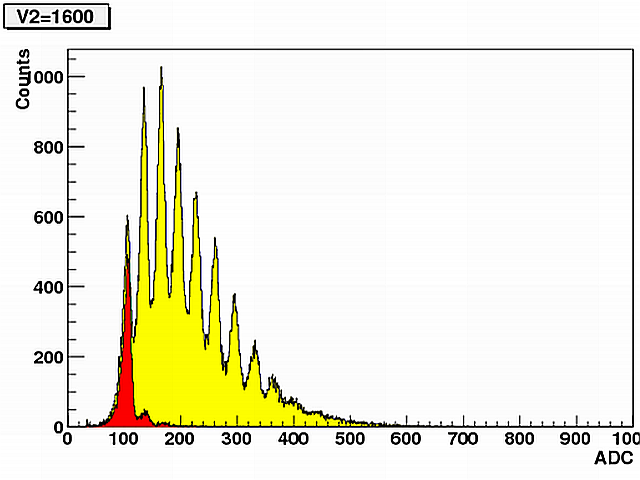}
\includegraphics[width=1.7in]{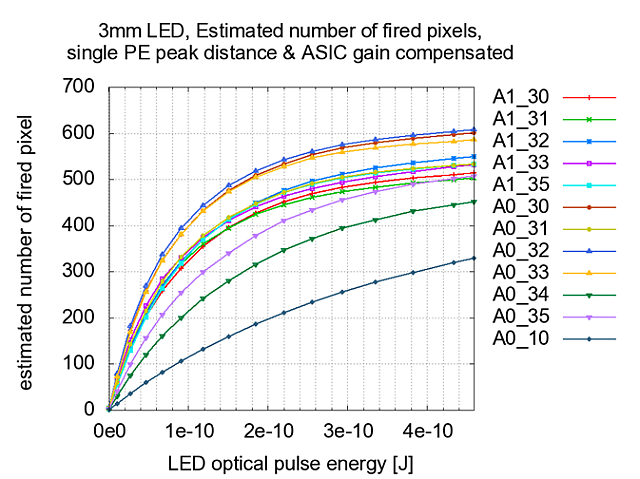}
\caption{QMB6 performance: (a) single photon spectrum measured at low light intensity 
(red color represents pedestal, no light); (b) saturation curves for all tiles illuminated by
 one UV-LED with pulse of 3.5~ns~width.}
\label{QMB6Performance}
\end{figure}

 For the second option of the SiPM calibration considered for the engineering prototype, 
 a dedicated LED calibration board \cite{Zalesak:EUDET200807} was developed at 
Institute of Physics in Prague. It consists of 6 Quasi-Resonant LED drivers. The board is controlled by the CANbus, 
or by a DIP switches in the Stand-alone mode.

  Peak LED optical power was measured as 0.4~nJ in a single pulse with Thorlabs power-meter PM100d with S130VC 
sensor and 3~mm UV LED. The beam test at DESY  \cite{Zalesak:EUDET201021}
in 2010 showed that a single LED is able to saturate 12~tiles at once with 
signal equivalent to 200~MIPs. 
The pulse length is fixed at 3.5~ns by the "L" and "C" components (Fig.~\ref{Quasidriver}(a))
parameters. 

Examples of the QMB6 performance could be seen in Fig.~\ref{QMB6Performance} 
displaying a single-photon spectrum of one illuminated tile channel at low light intensity
 and saturation curves for all measured channels, respectively.
 
The test in 4~T magnetic field \cite{Zalesak:EUDET200905} 
showed almost negligible influence to the QMB6 operation ($<$1\%).

\begin{figure}
\centering
\includegraphics[width=3.4in]{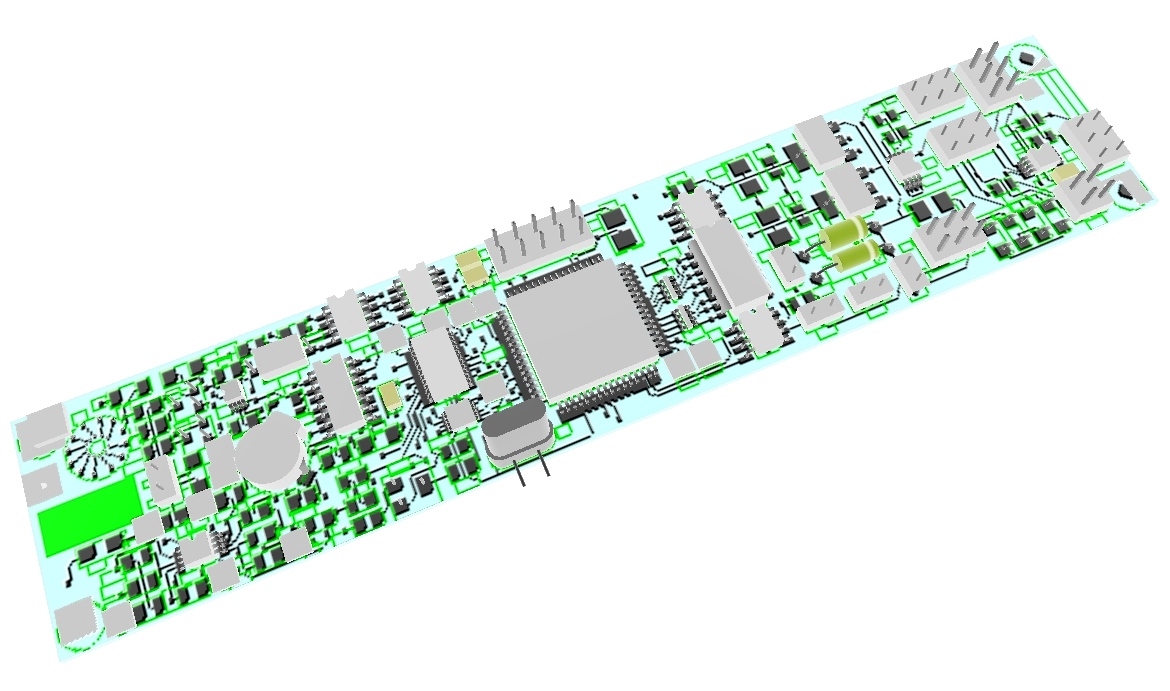}
\caption{3D view of the PCB layout of the QMB1}
\label{QMB1view}
\end{figure}

\begin{figure}
\centering
\includegraphics[width=3.4in]{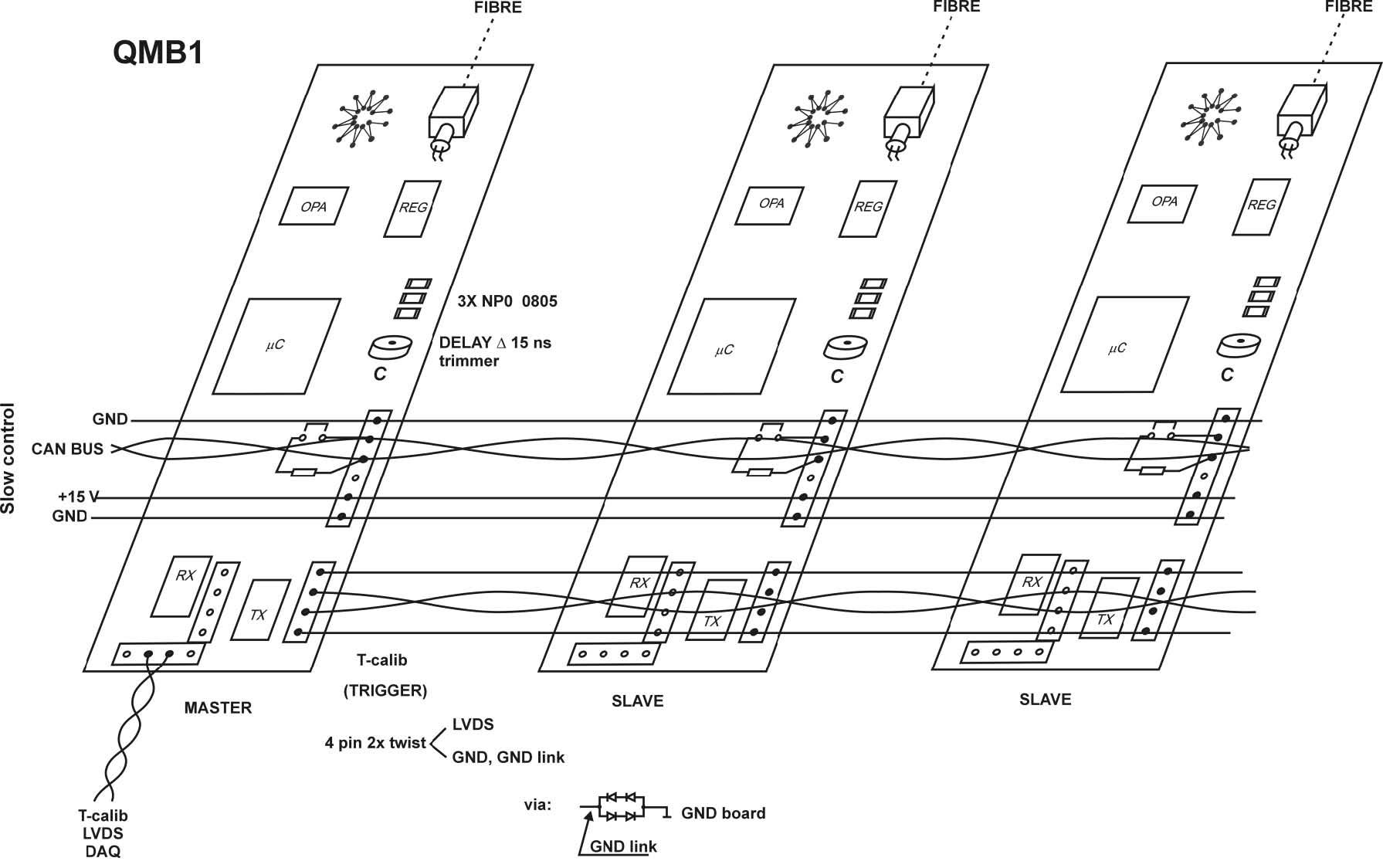}
\caption{Scheme of the QMB1}
\label{QMB1cheme}
\end{figure}

\subsection{1-ch Led Driver (QMB1)}

   Final 1-channel prototype (QMB1) is currently under development, in Fig.~\ref{QMB1view} we can see
three dimensional topology of designed QMB1 pcb. To fit the tile spacing the board is only 3~cm wide 
and 14~cm in depth. Each board consists only of one driver (for one UV or blue LED) but system is proposed 
as highly modular as one board behaves as a master and several others as slaves. The master board communicates 
with DAQ via CAN bus and it is also used as
a distributor of the LVDS trigger system as is pictured in scheme in Fig.~\ref{QMB1cheme}.
Pulse length is fixed to 5~ns to produce optical pulse. The light amplitude is smoothly variable 
(max~$\sim$1~A through a LED).
The goal is to illuminate 72~tiles by a single LED. This driver with optical fibers will be integrated to the 
newly built calorimeter prototype whose design can be seen in~Fig.~\ref{AhcalEngineer}.

\section{Light fiber distribution system\label{secFiber}}

The original concept, well proven by several years in operation in the beam tests, consists of a single 
LED sending light to a bunch of 19 fibers carried by an optical connector. 18 fibers are routed 
each to a single tile, 1 fiber was routed back to a PIN diode  for the feedback.

\begin{figure}
\centering
\includegraphics[width=1.3in]{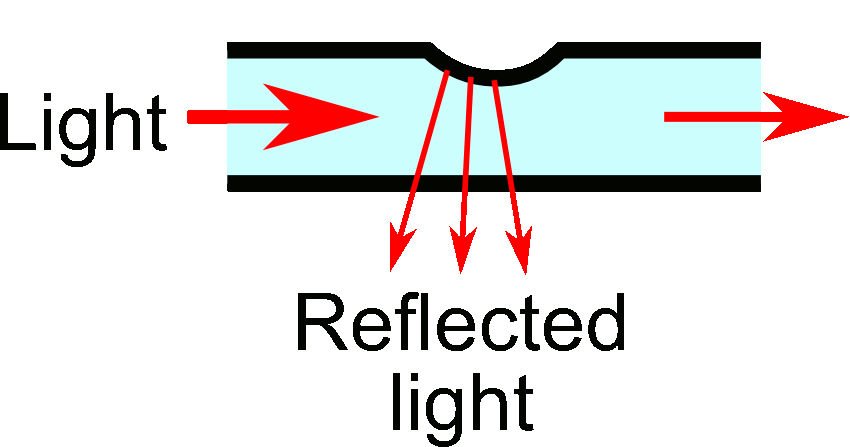}
\includegraphics[width=1.0in]{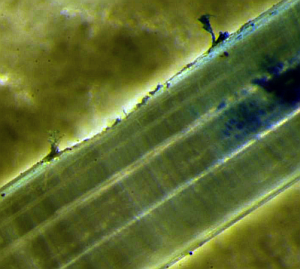}
\includegraphics[width=1.0in]{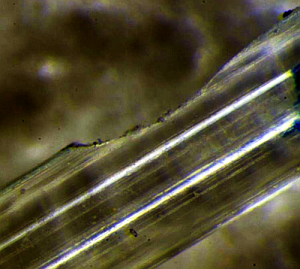}
\caption{(a) principle of the light emitted by the notch; (b) first notch, closest to the LED; (c) last notch, 
at the rear fiber end.}
\label{NotchedFiber}
\end{figure}

  In the next step we simplified the distribution of UV flashes to scintillator tiles. 
Instead of using one fiber for each tile, a series of cuts on a single optical fiber
 illuminates a row of tiles below the fiber.
The challenge is to make the light flashes equal with the same amount of light for each tile.
 
\begin{figure}
\centering
\includegraphics[width=3.5in]{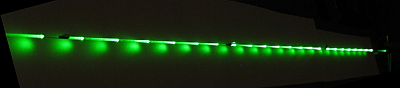}
\caption{ View of the optical fiber with notches. The fiber is illuminated by a 
continuous light to visualize the notches.}  
\label{GreenNotchFiber}
\end{figure}

  The specially ''notched'' plastic optical fiber has
spots along its length.
The notch is a special cut (Fig.~\ref{NotchedFiber}(a)) on the fiber's surface, which reflects light 
perpendicularly to the fiber path (Fig.~\ref{GreenNotchFiber}).
Depth of notches varies from beginning to the end of the fiber to guarantee uniform light 
output from each notch (Fig.~\ref{NotchedFiber}(b,c)).

\begin{figure}[!b]
\centering
\includegraphics[width=2.9in]{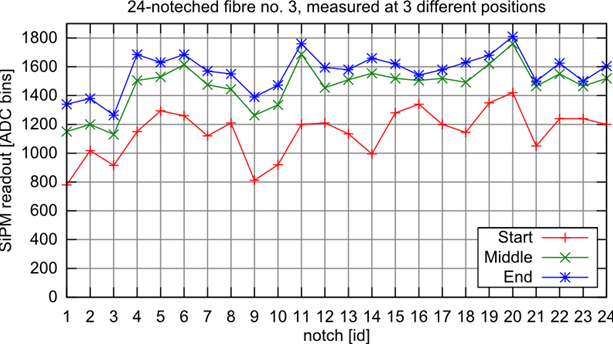}
\caption{Light output measurement as function of the position of a notch along fiber}
\label{ResultUniformity}
\end{figure}

Test of several  hand-made prototype fibers with different number of notches were performed. 
We achieved the spread of light better than 20\% on the fiber length 
of 210~cm with 72~notches~\cite{Zalesak:EUDET200807}. Other options,
24 and 12-notched fibers have been manufactured with $\pm$15\% and $\pm$10\% homogeneity,
respectively. 
An example of results on light output uniformity measurement with 24 notches is displayed
in Fig.~\ref{ResultUniformity}.
 
\begin{figure}
\centering
\includegraphics[width=2.675in]{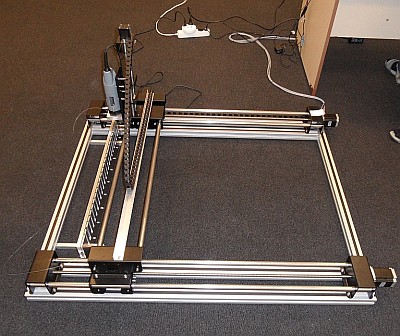}
\caption{Semi-automatic notch-fiber machine }
\label{SAM}
\end{figure}

\begin{figure}[!b]
\centering
\includegraphics[width=3.4in]{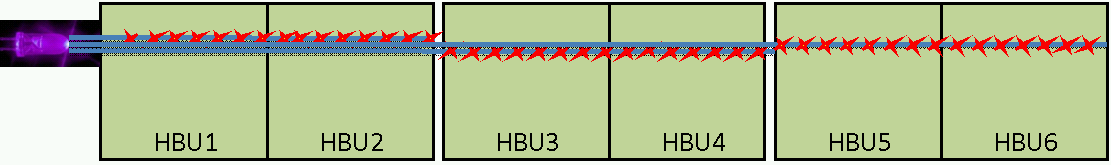}
\caption{Foreseen illumination of row with 72 tiles on 6 HBUs by 3 fibers with 24 notches each.}
\label{HBU3Fiber}
\end{figure}

  The optical fiber development continues with building a semi-automatic machine (Fig.~\ref{SAM})
for the notched fiber production.
New optical coupling to the bunch of 3 optical fibers is being developed. In the final prototype the light will be guided from single 
LED by 3 fibers having 24 notches each as is displayed in Fig.~\ref{HBU3Fiber}.
 Fibers will be routed on the component side of the HBU, where SMD components 
are not mounted. The light from the fiber will be guided to the tiles (sitting on the back side) through 
the 3 mm hole in the PCB of HBU.

\section*{Acknowledgment}

The author gratefully thanks to his colleagues from the Institute of Physics in Prague: 
Jaroslav Cvach, Milan Janata, Jiri Kvasnicka, Denis Lednicky, Ivo Polak, Jan Smolik 
for very valuable contributions to the results presented in this paper. 
Special thanks go to  Mark Terwort and Mathias Reinecke (DESY) 
for their help and realization the measurements needed for this work.


\end{document}